\documentclass[11pt]{article}
\usepackage{graphics}
\usepackage{epsfig}
\usepackage{amssymb}

\def\scrip{{\mathcal I}^-}
\def\scrif{{\mathcal I}^+}

\date{\empty}
\begin{document}
\title{Quantum Character of Black Holes}

\author{Adam D. Helfer\\ 
{\small Department of Mathematics, University of Missouri,
Columbia, MO 65211}}
\maketitle

\begin{abstract}  
Black holes are extreme manifestations of general relativity, so one might hope
that exotic quantum effects would be amplified in their vicinities, perhaps
providing clues to quantum gravity.  The commonly accepted treatment of 
quantum corrections to the physics around the holes, however, has provided only
limited encouragement of this hope. The predicted corrections have been minor
(for macroscopic holes):  weak fluxes of low-energy thermal radiation which
hardly disturb the classical structures of the holes. Here, I argue that this
accepted treatment must be substantially revised.  I show that when
interactions among fields are taken into account (they were 
largely neglected in the
earlier work) the picture that is drawn is very different.  Not only low-energy
radiation but also ultra-energetic quanta are produced in the gravitationally
collapsing region.  The energies of these quanta grow exponentially quickly, so
that by the time the hole can be said to have formed, they have passed the
Planck scale, at which quantum gravity must become dominant.  The vicinities of
black holes are windows on quantum gravity.

\end{abstract}             

\begin{center}
\end{center}

\begin{center}
\end{center}


\section{Introduction}
\label{sec:intro}

Quantum theory and general relativity are at once practically compatible and
theoretically irreconcilable.   While there is a profound difficulty in melding
the theories, it is not apparent at ordinary scales.  It is expected to become
pronounced rather in extreme circumstances, in the Planck regime.

The difficulty in reconciling the theories derives from a fundamental
incompatibility in their hypotheses. According to quantum theory, the position
and momentum of an object can never be known simultaneously with unlimited
accuracy; but according to general relativity, we should always be able to
imagine space--time populated with locally inertial observers who can make
physical measurements  to arbitrary accuracy and without disturbing the
underlying geometry of space--time.    For measurements at ordinary scales,
this conflict is utterly insignificant.  But for measurements at very fine
scales we must use, according to quantum theory, particles with high momenta. 
And at some point --- the Planck threshold --- the momenta become so high that
their gravitational effects interfere with the measurements being attempted. 
At such scales it becomes impossible to give operational significance to
ordinary notions of space, time, momentum and energy \cite{DeWitt:1962}. The
occurrence of Planckian or ``trans-Planckian'' (past the Planck scale)
quantities in a conventional physical theory is a sign it has been applied
beyond its realm of credibility.

I shall show here that, when space--time gravitationally collapses to form a
black hole, quantum theory predicts the production of particles at, and the
scattering of particles to, Planckian energies.  This means the theory predicts
its own inapplicability, and the onset of an essentially quantum-gravitational
regime, in the vicinity of a black hole.  The affected region of space--time
extends outwards from the horizon a distance comparable to the size of the
hole. These results do not allow us to say precisely what the
quantum-gravitational effects in the vicinities of black holes should be ---
for that,  we would need a theory of quantum gravity.  But we do have an
unequivocal prediction that the quantum-gravitational effects should be
significant in a substantial volume of space--time.   Therefore observations of
particles from the vicinities of black holes should be able to provide us with
clues about quantum gravity.

The picture that is drawn here is very different from the accepted one (due to
Hawking's pioneering analysis \cite{Hawking:1974,Hawking:1975,Helfer:2003}), 
in which black holes are essentially classical objects and quantum corrections
to their structure are minor.  The reason for this discrepancy is that concerns
about the accepted theory, which had generally been dismissed as niggling, turn
out to be very much on point and to alter the picture significantly.
Trans-Planckian ``virtual'' effects (which had been considered an awkward
peripheral feature of the conventional theory) and interactions between quantum
fields (which had been largely neglected\footnote{An important attempt to
incorporate interactions is due to Gibbons and Perry~\cite{GP:1976}.  The
relation of their approach to the present work is discussed in the appendix.})
combine to produce real Planckian effects.

A rather different argument, but reaching parallel conclusions, has been given
elsewhere~\cite{Helfer:2004}.

Here is the plan of the paper.  Sections~\ref{BH} and~\ref{FHR} review the
relevant structure of a gravitationally collapsing space--time and of the
Hawking process.  They emphasize the origin of the Hawking quanta in ultra-high
frequency vacuum fluctuations in the distant past; as the associated virtual
quanta propagate through the gravitationally collapsing region they are
distorted and red-shifted by an exponentially growing factor.  Since they
emerge with frequencies of the order of the characteristic Hawking frequency,
the precursors' frequencies grow exponentially.  

Section~\ref{EI} contains the computation of the effect of interactions, in the
case of quantum electrodynamics.  I show there that there is a non-trivial
first-order amplitude to produce  electron--positron pairs, at the
exponentially growing energies associated with the Hawking quanta's
precursors.  One can think of the process as being due to the
quantum-electrodynamic dressing of the vacuum by virtual triples, each
consisting of a photon and an electron--positron pair.  For an ultra-energetic
such triple, the photon may pass through the collapse region and be distorted
to a Hawking quantum which can no longer recombine with the electron--positron
pair to produce the proper dressing of the vacuum.  The result is a real
Hawking photon together with a real ultra-energetic electron--positron pair.

Section~\ref{CP} discusses the consequences of the analysis.  The exponential
growth in the energy scales means that one very quickly passes the regime in
which reliable theoretical computations of quantum-field-theoretic processes
are possible, and in a few dozen $e$-foldings has arrived at the Planck regime,
where the entire theory breaks down.  Thus our main conclusion is that current
theory is unable to make reliable predictions about quantum physics in the
neighborhood of a black hole.  Indeed, we suggest that observations of the
vicinities of black holes may be able to provide us with evidence of the
behavior of quantum field theories at ultra-high energies and the character of
quantum gravity.

\textit{Conventions and terminology.}
We shall use natural units, so $c$, $G$ and $\hbar$ are unity.    We shall need
to discuss processes with  energies approaching, but not at, the Planck energy;
we shall refer to these as ultra-energetic. 
For simplicity, in most of the paper we consider only an isolated uncharged
spherically symmetric black hole;\footnote{In subsection~\ref{dis} we shall
explain that the results here hold too for holes with angular momentum and
charge.} its mass will be $M$.
Conventions for quantum fields are those of Schweber~\cite{Schweber:1961}.

\section{Black Holes}
\label{BH}

While the event horizon of a black hole is best known as a ``point of no
return,'' it is a different property which is most important both in Hawking's
analysis and here.  This is that light rays (and other fields) passing close to
the event horizon are red-shifted, the shifts increasing exponentially as later
and later rays are considered.

It is convenient to introduce a diagram in which the past and future limits of
the light rays can be represented.  In order to do this, we distort the scales
so that these limits appear at finite positions on the diagram.  It is possible
to do this while retaining an accurate representation of causal relations.

This is done in Fig.~1, which suppresses the angular variables and thus
represents the radial and temporal degrees of freedom.  The spatial origin is
at the left-hand edge, and radial light rays are at $45^\circ$.   The event
horizon is the dashed line, and the points at and above this form the black 
hole itself.  

\begin{figure}[t]
\epsfxsize=2in
\epsfbox{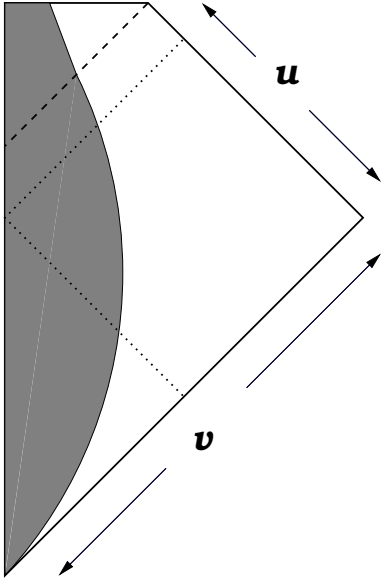}
\caption{A diagram of a black-hole space--time, suppressing angular
variables.  (The left-hand edge is the spatial origin.) Time increases
upwards, and lines at $45^\circ$ represent the paths of radial light
rays.   The scale has been distorted so that the entire space--time, and some
ideal points at infinity, can be represented.  The region occupied by the
collapsing matter is shaded.  The event horizon ${\cal H}$ is the dashed line; 
the black hole itself is the set of points at and above this. The dotted line
represents a radial light ray beginning at a point on $\scrip$ (an ideal set in
the past, coordinatized by the advanced time $v$), moving radially inwards and
passing through the spatial origin (where, in the diagram, it appears to
reflect from the left-hand edge), and escaping to a point on the ideal set
$\scrif$ (coordinatized by $u$).  The sets $\scrip$ and $\scrif$ are really
unbounded, but, because of the distortion of scales appear in the diagram to be
finite.}
\end{figure}

The future limits of radial light rays are on the upper line at $45^\circ$,
coordinatized by the ``retarded time'' $u$, and denoted collectively $\scrif$.
(Similarly, the past limits are coordinatized by ``advanced time'' $v$ and
denoted $\scrip$.) One should think of the neighborhood of $\scrif$ as the
locations of observers far from the collapsing object who look back into the
space--time with $u$ their appropriate measure of time. (The neighborhood of
$\scrip$ is the set of locations of distant actors who might send signals into
the space--time.) These external observers are, by definition, everywhere below
the event horizon.  This means that later and later external observers appear,
in Fig.~1, to crowd the upper-right corner.  This is only the effect of the
distortion of scales, however; the corresponding portion of space--time is in
reality unbounded. Similarly, the set $\scrif$ is really unbounded in the past
(that is, towards the right-hand corner in Fig.~1), and the set $\scrip$ is
likewise really unbounded.

An observer at a retarded time $u$ near $\scrif$, looking inwards, would see a
radial light ray.  Tracing this ray backwards in time, it would move inwards to
the spatial origin, pass through this (appearing, in the diagram, to reflect
from the left-hand edge), and emerge finally at an advanced time $v(u)$ on
$\scrip$ (dotted line in Fig.~1).  The periods $dv$ and $du$ of a light ray at
$\scrip$ and $\scrif$ are related by $dv=v'(u)\, du$, so $v'(u)$ is the
red-shift suffered by the ray.

The red-shift factor $v'(u)$ plays a key role in quantum physics around black
holes.  It can be shown to have the asymptotic form 
\begin{equation}
v'(u)\simeq \exp
-u/(4M)\quad\hbox{as}\quad u\to +\infty\, ,\label{eq:asympt}
\end{equation}
where $M$ is the mass of the hole \cite{MTW:1973,Helfer:2001}.  This
exponential decay drives $v'(u)$ to zero very quickly.  (For example, for a
solar-mass object, the $e$-folding time is $\simeq 2\times 10^{-5}$ s.)
Although observers in the exterior of the black hole (by definition) never
cross the event horizon and so never, strictly speaking, see the hole, the
nominal time of formation of the hole may be taken as the point where $v'(u)$
has decreased to become practically indistinguishable from zero.

\section{Fluctuations, 
Hawking Radiation\\ and the Trans-Planckian Problem}
\label{FHR}

Because the uncertainty principle forbids measuring all degrees of freedom of a
quantum field simultaneously, one can never speak of a state of identically
zero field.  This means that even in a space--time unaffected by gravity the
vacuum is populated by quantum fluctuations in the field, sometimes called
zero-point fluctuations.  While they do not contribute any real particles to
the state, they can have significant indirect effects.

The lowest-order vacuum fluctuations in the absence of gravity are represented
in Feynman-type  diagrams by closed loops (Fig.~2).  These loops suggest
particles coming into existence and annihilating themselves.  While this is not
wholly correct --- the use of the term ``particle'' in this context is
oversimplistic, and the loops do not have any distinguished points at which one
can say creation or annihilation occurs --- it will not be necessary for us to
refine this.  In general, intermediate ``particles'' are called ``virtual''.
(While its literal significance will not be needed here, Fig.~2 corresponds to
the effect of zero-point energy  on the propagation of the field.  This effect
can be consistently subtracted when there is no explicit time-dependence, and
in those cases diagrams like Fig.~2 can  be neglected and are rarely drawn.)

\begin{figure}
\epsfxsize=1in
\epsfbox{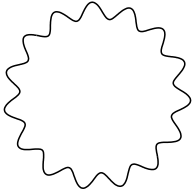}
\caption{A zero-point fluctuation's contribution to the evolution,
represented by a closed virtual photon loop. (Wavy lines represent photons.)
}
\end{figure}

In principle, according to quantum field theory, vacuum fluctuations occur for
all field modes, including those of arbitrarily high frequencies, corresponding
to virtual particles of arbitrarily high energies.    This would be a problem
if the effects of arbitrarily high-energy fluctuations had always to be
considered, because quantum field theory itself should break down at the Planck
scale and quantum gravity should take over.  However, in realistic
theories, for any given process, there is an energy scale beyond which
the fluctuations decouple and can be ignored.  As long as this energy
is below the Planck scale there is no trouble.
We shall see, however, that Hawking's model requires the use of virtual
particles at arbitrarily high energies, and so its reliance on conventional
quantum field theory is inappropriate.

When the quantum fields propagate through a time-dependent region of
space--time, they are distorted and the balance of the vacuum fluctuations is
upset.  This is what, in Hawking's analysis, gives rise to thermal radiation. 
We can represent this by drawing Feynman-type diagrams on top of space--time
diagrams, as in Fig.~3.  In these diagrams, one can think of the effect of the
passage through space--time as opening some of the vacuum fluctuation loops. 
The results are real particles emitted, and those which escape to $\scrif$ are
the Hawking radiation.  

\begin{figure}[t]
\epsfxsize=2in
\epsfbox{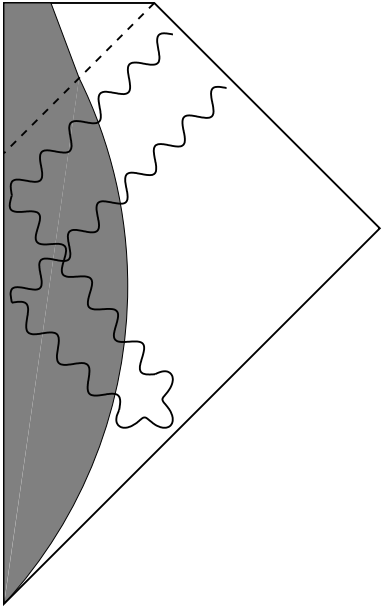}
\caption{One can think of propagation through the 
gravitationally collapsing space--time as opening some of the 
vacuum fluctuation loops, resulting in real particle production.  
Although, for fixed end-points, one could draw vacuum fluctuation
arcs occupying any portion of space--time, the dominant contributions
to the Hawking process come from ones like that shown here.
Ultra-high frequency fluctuations in the past propagate through the
collapsing space--time, where they are both red-shifted and distorted
in a way which produces a low flux of real low-energy photons.
(The red-shift and distortion are not represented in this diagram.)
Thus while the outgoing arms of the diagram represent real particles, the
portion deep within space--time has a more schematic significance.
}
\end{figure}

Although virtual particles can propagate throughout space--time, only certain
vacuum fluctuations propagated in particular ways will be distorted in the
correct manner to give rise to significant real particle production.  The most
important contributions to the Hawking process come from modes like those shown
in Fig.~3. These originate as vacuum fluctuations in the past, propagate
inwards, pass through the origin and then move outwards to $\scrif$ just before
the event horizon.  Hawking's prediction of thermal radiation, although not
originally expressed diagrammatically, is equivalent to the evaluation of 
diagrams like Fig.~3, and Hawking's remarkable insight was (again, recast in
diagrammatic form) that the main contributions would come from the modes shown
there.

The red-shift of the fluctuations as they propagate through space--time and
reify is crucial.  The predicted frequencies of the Hawking particles are $\sim
\omega _{\rm H}=1/(8\pi M)$.  However, these frequencies have been red-shifted
from those of their vacuum-fluctuation precursors.  The precursors' frequencies
must have been $\sim \omega _{\rm H}/v'(u)\simeq \omega _{\rm H}\exp +u/(4M)$. 
So the precursors --- which should be thought of as the portion of Fig.~3
around and including the semicircular portion at the lower right --- have
frequencies which grow exponentially and rapidly reach the Planck scale.   Then
quantum-gravitational effects must take over.   

This is the ``trans-Planckian problem'' \cite{Gibbons:1977,Helfer:2003}.   The
prediction of thermal radiation relies on assuming that conventional physics
applies to vacuum fluctuations at exponentially increasing energy scales.  The
scales rapidly pass the Planck threshold, at which quantum-gravitational
effects must supplant conventional physics. The conventional  analysis
\cite{Hawking:1974,Hawking:1975} ignores this, and uses vacuum fluctuations at
the exponentially increasing, trans-Planckian, scales. For us, however, the
ultra-energetic portion of Fig.~3 will be of central
significance.\footnote{There is one approach to the trans-Planckian problem 
which has been much pursued in recent years and is worth mentioning here.  
Unruh~\cite{Unruh:1995}, Corley and Jacobson~\cite{CJ:1996} and others (see 
ref.~\cite{Helfer:2003} for more references and discussion) aim to resolve the
problem by substituting non-standard rules for the propagations of the fields. 
Their goal is to ``insulate'' Hawking's  result from trans-Planckian physics,
and this is done by radically altering the actual mechanism that is used to get
that result. At the moment, work on these approaches is ad-hoc and preliminary,
and the extent to which they can be said to eliminate the trans-Planckian
problem is not entirely clear.  If, however, one of these non-standard 
approaches were to prove correct (that is, could be developed into a  theory
which turned out to be the way the world works), then the present  analysis,
which relies heavily on the original Hawking model, would have to be
reconsidered.}

\section{Effects of Interactions}
\label{EI}

Hawking's analysis rested on several assumptions,  one of which was
that interactions between quantum fields could be neglected.  To my knowledge,
there has been no attempt to trace through the essential logic of the analysis
while taking interactions into account. And yet this is surely worthwhile, for,
while interacting quantum field theories are admittedly difficult, they are
well understood (at least perturbatively).  

We shall consider for definiteness quantum electrodynamics, the theory of
photons, electrons and positrons, but it will be apparent that the main ideas
are more general.  Because the coupling is weak, in most familiar situations,
it is legitimate to treat this theory as a perturbation of a ``bare'' theory of
noninteracting photons and charged particles.  The interaction is determined by
the vertex in Fig.~4, which may represent scattering of a charged particle by a
photon, or a pair of charged particles converting to or from a photon.

\begin{figure}
\epsfxsize=1in
\epsfbox{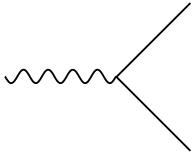}
\caption{The interaction between the electromagnetic field (wavy line) and
charged particles (solid lines).  A photon can scatter a charged particle, or,
in suitable circumstances the electromagnetic field can create pairs of
particles, or pairs of particles can annihilate with a release of 
electromagnetic quanta.
}
\end{figure}

We shall see that the perturbation theory in the presence of
gravitational collapse is qualitatively different from that in
Minkowski space, and that this difference becomes apparent at first
order (in the electric charge).  At this order, we might expect to
draw diagrams like Fig.~5, which (intuitively, at least) would
represent the production of a pair of charged particles,
together with a Hawking-type photon, by a vacuum fluctuation.
A diagram like this would vanish in Minkowski space, by conservation
of energy--momentum, but in a gravitationally collapsing space--time
some of the energy--momentum may be exchanged with the incipient black
hole.  We shall see that this diagram has a non-zero value, and in
fact represents the pair production of ultra-energetic charged
particles.

\begin{figure}
\epsfxsize=2in
\epsfbox{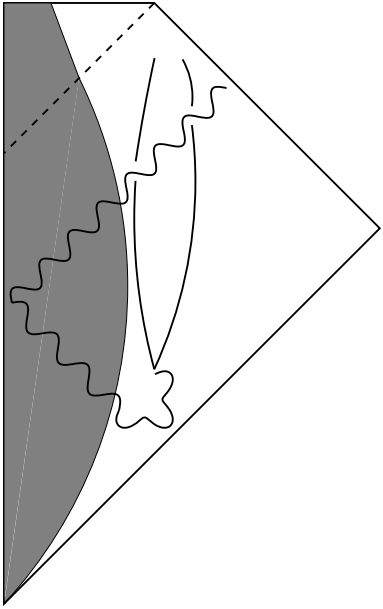}
\caption{An ultra-high energy vacuum electromagnetic fluctuation (wavy
line) can produce an ultra-energetic pair of charged particles
(solid lines).  (The bending-back on itself of the photon line is meant to
suggest its origin in vacuum fluctuations, but has no
literal significance.)
}
\end{figure}

We shall now make this precise.  However, we shall do so in terms of
conventional operator-theoretic methods, rather than Feynman-type diagrams. 
(There is no standard definition of such diagrams in curved space--time, and
making one and carefully connecting it with conventional field theory would be
lengthy; also the operator-theoretic treatment is better suited to some aspects
of the discussion.)

It should be emphasized that our goal here is to clearly establish the existence
of real ultra-energetic effects.  It is not to make quantitatively accurate
estimates of them in very general circumstances, or even to discuss the variety
of different possible effects.  While such estimates and
discussion are of interest in themselves, they are not of primary interest in
the present work.  This is because, once we know that real ultra-energetic
effects exist, the exponentially increasing energies quickly move the effects
beyond those which can be reliably computed.  The main point of showing that the
effects are real is to show that the theory itself breaks down.

We shall therefore
choose to examine a particular situation in which the effects
are revealed with a minimum of computational and interpretational
technicalities.  This will be the main point of subsection~\ref{comp}, below.
First, however, we must understand what the physical vacuum is in the
interacting case.

\subsection{Dressing the Vacuum}

It will be crucial to understand how the vacuum is altered when the interaction
is ``switched on.''  The vacuum in question is the in-vacuum, defined as the
state of lowest energy in the distant past.  In this regime, we shall assume
that space--time, at least exterior to the collapsing body, is to good
approximation stationary (in fact, in the case we are considering, spherical
symmetry and vacuum exterior force it to be exactly Schwarzschild in this
region).  Then we may write the Hamiltonian in the past as $H_{\rm p}=H_{\rm
b}+H_{\rm int}$, where $H_{\rm b}$  is the ``bare,'' noninteracting, term and
$H_{\rm int}$ is the interaction.  According to standard perturbation theory,
the bare in-vacuum $|0_{\rm b}\rangle$ is perturbed to first order to a
``dressed'' vacuum
\begin{equation}
|0_{\rm d}\rangle =\left( 1-H_{\rm b}^{-1} H_{\rm int}\right) |0_{\rm b}\rangle
  \, .
\end{equation}
The interaction Hamiltonian is $H_{\rm int}=-e\int _\Sigma 
\tilde\psi \gamma ^a\Phi _a\psi $, where the integration is over the initial
surface in question, and $\Phi$ and $\psi$ are the electromagnetic and
charged-particle field operators.  Each of these operators is a sum of (bare)
creation and annihilation terms.  Combinations involving annihilation terms will
not contribute to the perturbation of the vacuum, but those involving creation
terms from $\tilde\psi$, $\Phi$ and $\psi$ will.  This is one way of
understanding how the interaction populates
the vacuum with virtual quanta.

\subsection{Charged-Pair Production}\label{comp}

While, as emphasized above, our computation will be in
operator-theoretic terms and will not rely on diagrams, the diagrams
do provide a useful intuitive picture, and we shall begin by
explaining the connection between the quantity to be computed and them.

We shall compute an amplitude which can be thought of as corresponding to the
diagram in Fig.~5.  Our point of view is that this diagram is a modification of
the conventional Hawking process, where one branch of the electromagnetic line
has been converted to an electron--positron pair.  Thus the diagram represents
the {\em destruction} of a Hawking electromagnetic quantum (relative to the
in-vacuum --- recall that the in-vacuum is the state which has Hawking quanta
in the out-region) and the creation of a charged-particle pair.  We may write
the amplitude as
\begin{equation}\label{expval}
\langle 0_{\rm d}|{\tilde\psi}^+_{\rm f}(k')\Phi
^-_{\rm f}(k)\psi ^+_{\rm f}(k'')|0_{\rm d}\rangle\, ,
\end{equation}
where ${\tilde\psi}^+_{\rm f}$, $\Phi ^-_{\rm f}$, $\psi ^+_{\rm f}$ are
annihilation  and creation operators in the distant future, and $k'$, $k$,
$k''$ are mode labels.  

In order to simplify the analysis, we shall choose to examine the
amplitude~(\ref{expval}) only for certain modes.   Most importantly, the
charged-particle modes will be restricted so that to zeroth order they
correspond to wavepackets propagating everywhere throughout the vacuum region. 
This means that to zeroth order the charged-particle modes are unaffected by
the dynamic, collapse, portion of the space--time. (In particular, we do not
then have to consider Hawking-production of charged particles, for
Hawking-produced particles appear to emanate from the collapsing region.) It
will also be convenient to consider only modes corresponding to propagations
everywhere far from the strong-field region.  This is not really essential, but
allows us to use a conventional momentum-space representation of the field
operators.

With these restrictions on the modes to be examined,  the
amplitude~(\ref{expval})  vanishes to zeroth order, because the
charged-particle field $\psi ^+_{\rm f}(k'')$ annihilates the quantum state in
this order.  We shall see, however, that there are non-trivial first-order
effects.

There are two sorts of first-order contributions to the 
amplitude~(\ref{expval}):   those where the first-order correction is to one of
the operators; and those where it is to the state vector.  As it turns out,
those due to corrections to the operators do not produce very significant
effects.  

(The reason for this is essentially the same as in Minkowski space.
The corrections due to the perturbations of the operators are
\begin{equation}
\langle 0_{\rm b}| \left({\tilde\psi}^+_{{\rm f}b}\Phi
^-_{{\rm f}b}\Delta\psi ^+_{\rm f}+{\tilde\psi}^+_{{\rm f}b}\Delta\Phi
^-_{\rm f}\psi ^+_{{\rm f}b}+\Delta {\tilde\psi}^+_{\rm f}\Phi
^-_{{\rm f}b}\psi ^+_{{\rm f}b} \right) |0_{\rm b}\rangle\, ,
\end{equation}
where in each case the subscript ``b'' denotes the bare term and the
prefix $\Delta$ the first-order correction.  Of these, the latter two
vanish since the charged-particle annihilation operator is applied to
the vacuum; we are left with
$\langle 0_{\rm b}| {\tilde\psi}^+_{{\rm f}b}\Phi
^-_{{\rm f}b}\Delta\psi ^+_{\rm f}
|0_{\rm b}\rangle$.
This leads to an interaction integral of the form
\begin{equation}
  \int \tilde v \gamma ^aA_a u\, d\tau\, ,
\end{equation}
where $\tilde v$, $A_a$ and $u$ are the mode functions for the fields,
determined from their data in the future and propagated through
space--time by the free field equations,
and the integration extends over the space--time volume in question.
If the mode functions correspond to particle wavepackets sufficiently
well-localized that the volume in which there is a significant
interaction is small compared to the space--time curvature, then in
this volume the interaction must conserve energy--momentum. 
In this interaction volume, both the electromagnetic and
electron--positron mode functions can be thought of as on-shell modes
in Minkowski space.  The electron--positron mode functions will
correspond to a superposition of terms, each with timelike
future-pointing energy--momentum.  The electromagnetic mode functions,
on the other hand, will be a superposition of terms each with null
energy--momentum.  It will therefore be impossible for
energy--momentum to be conserved:  the interaction integral will
vanish due to destructive interference.)

The interesting first-order contributions to the quantity~(\ref{expval}) come
from the perturbation of the vacuum.  In principle, there are two of these, one
from the perturbation of the bra and the other from the perturbation of the
ket, but the first of these vanishes as in it annihilation operators are
applied directly to the bare vacuum.  We have, therefore, 
\begin{equation}
\langle 0_{\rm d}|{\tilde\psi}^+_{\rm f}\Phi
^-_{\rm f}\psi ^+_{\rm f}|0_{\rm d}\rangle =
-\langle 0_{\rm b}|{\tilde\psi}^+_{\rm f}\Phi
^-_{\rm f}\psi ^+_{\rm f}H_{\rm b}^{-1}H_{\rm int}|0_{\rm b}\rangle\,
,
\end{equation}
to the approximation required, where the field operators may be taken
to be bare.

We should, strictly speaking, do the details of the computation with
wave-packets to represent the mode functions determining the fields.  However,
as is conventional, we shall work directly with the momentum-space
representations of the fields, and understand that the computation really only
has meaning when averaged over appropriate wave-packets (corresponding to the
assumptions we have made about the modes we consider).

We shall therefore take the mode labels $k'$, $k$, $k''$ to refer to
three-momenta in the out-region.  (We will supplement these with polarizations
shortly.) 
Since to zeroth order the spinor field modes of interest for us are unaffected
by the space--time curvature, it is not necessary to distinguish, to this
order, the momentum-space decomposition of the spinor fields in the future from
that in the past.  However, for the electromagnetic field this distinction is
essential.  We shall put
\begin{equation}
\Phi ^+_{\rm f}(k)=\int \left[ \alpha (k,l)\Phi ^+_{\rm p}(l) +\beta (k,l)\Phi
^-_{\rm p}(l)\right]\, d^3l\, ,
\end{equation}
where $\alpha (k,l)$, $\beta (k,l)$ are called the Bogoliubov coefficients of 
the (zeroth order) scattering.  

We may now straightforwardly compute the amplitude~(\ref{expval}):
\begin{eqnarray}
\lefteqn{\langle 0_{\rm d}|{\tilde\psi}^+_{\rm f}(k')\Phi
^-_{\rm f}(k)\psi ^+_{\rm f}(k'')|0_{\rm d}\rangle}\nonumber\\
 &=&
-\int d^3l\langle 0_{\rm b}|{\tilde\psi}^+(k') \overline{\beta (k,l)}
\Phi ^+_{\rm p}
(l)
\psi ^+(k'')H_{\rm b}^{-1}H_{\rm int}|0_{\rm b}\rangle\nonumber\\
 &=&
-\int d^3l\left[ E(k')+\| l\|+ E(k'')\right] ^{-1}\nonumber\\ &&\times
   \langle 0_{\rm b}|{\tilde\psi}^+(k') \overline{\beta (k,l)}\Phi ^+_{\rm p}
(l)
\psi ^+(k'')H_{\rm int}|0_{\rm b}\rangle\, ,
\end{eqnarray}
where $E(k')=\sqrt{\| k'\|^2 +m^2}$ is the energy of a particle of
mass $m$ and momentum $k'$.  
We have
$H_{\rm int}=-e\int _\Sigma\tilde\psi\gamma ^a\Phi _a\psi $, and when we compute
the expectation
$\langle 0_{\rm b}|{\tilde\psi}^+(k') \Phi ^+_{\rm p} (l)
\psi ^+(k'')H_{\rm int}|0_{\rm b}\rangle$, the annihilation operators applied to
the fields in $H_{\rm int}$ simply produce the corresponding mode functions. 
With standard continuum normalizations for these
(the operator
${\tilde\psi}^+(k')$ corresponding to the mode $(2\pi )^{-3/2}
(m/E(k'))^{1/2} e^{-ik'x}
v(k')$, the operator $\Phi ^+(k)$ to $(2\pi )^{-3/2} \| 2k\| ^{-1/2}e^{-ikx}
\epsilon (k)$ and $\psi ^+(k'')$ to $(2\pi )^{-3/2}
(m/E(k''))^{1/2} e^{-ik''x} {\tilde v}(k'')$, where we have written the
polarizations $v(k')$, $\epsilon (k)$, ${\tilde v}(k'')$ explicitly)
we find, after a short calculation,
\begin{eqnarray}\label{evamp}
\lefteqn{\langle 0_{\rm d}|{\tilde\psi}^+_{\rm f}(k')\Phi
^-_{\rm f}(k)\psi ^+_{\rm f}(k'')|0_{\rm d}\rangle
=e m\pi ^{-3/2} \| 4(k'+k'')\| ^{-1/2}}\nonumber\\
&&\times (E(k')E(k''))^{-1/2}
\left[ E(k') +\| k'+k''\| +E(k'')\right]
^{-1} \nonumber\\
&&\times\tilde{v} (k'')\gamma^b
\overline{\beta _{ab}(k,-k'-k'')}\epsilon ^a(k) v(k') \, ,
\end{eqnarray}
where we have written the index structure on the Bogoliubov coefficient 
and the photon polarization explicitly.

\subsection{Interpretation}

The interpretation of the evaluated amplitude~(\ref{evamp})  turns on the
behavior of the Bogoliubov coefficient $\beta (k,l)$.  This coefficient was
evaluated by Hawking; indeed, its evaluation was the main goal of his work, for
the probability of producing Hawking photons in mode $k$ is $\sim d^3k
\int |\beta
(k,l)|^2 d^3l$.\footnote{In fact, Hawking used a decomposition in spherical
harmonics rather than the momentum-space one here.  However, it is easy to
interconvert between the two.}

Hawking found that the quantities were controlled by the characteristic 
frequency scale $\omega _{\rm H}=(8\pi M)^{-1}$ (where $M$ is the mass of the
collapsing object).  Associated with this is a characteristic period $\omega
_{\rm H}^{-1}$.  In any epoch (interval of late time a few characteristic
periods in duration), the main contributions to $\beta (k,l)$ come from
wavenumbers $k$ of magnitude $\sim \omega _{\rm H}$ but with $l$ the
corresponding, blue-shifted, precursor.  Thus the magnitude of $l$ is $\sim
\omega _{\rm H}\exp +u/(4M)$ and it is directed inwards.

Here, we have $l=-k'-k''$ with $k'$, $k''$ the momenta of the charged
particles. Thus we have ultra-high momenta $k'+k''$ directed {\em outward}. 
This is our amplitude to produce ultra-energetic pairs. The corresponding
probability, being proportional to $|\beta (k,l)|^2$ (modulo polarization
effects), can be thought of as representing the conversion of a fraction of
Hawking quanta into these pairs.

It is natural to ask what the implications of this result are for late
times $u$. 
However, once the momenta $k'$, $k''$ have become ultra-relativistic
(that
is, $\| k'\| ,\| k''\|\gg m$), first-order perturbation theory is no longer
valid; one must take higher-order corrections into account.  It is beyond the
range of our present computational abilities to make accurate evaluations of the
amplitudes when $u$ increases very much.\footnote{There are essentially two
competing effects as $u$ increases.  On the one hand, the blue-shifting of the
modes 
goes on in a region of
space--time which is itself being exponentially compressed; this tends to
decrease the amplitudes exponentially.  On the other, as the energies increase
one has more possible processes contributing to a given amplitude; estimates
of this effect are speculative.}
Indeed, the point of view we shall
advocate is that the computation here shows that non-trivial
ultra-energetic effects are possible, but it is unrealistic at present to expect
to be able to make quantitative theoretic predictions of them.  Rather, it may
be that we shall learn experimentally the physics of ultra-energetic quantum
fields by observing signals from the vicinities of black holes.

\subsection{Discussion}
\label{dis}

There is no wholly satisfactory brief account of the physics
underlying this ultra-energetic pair production, because of the interpretational
difficulties present in any interacting quantum field theory.  However, the
effect is important enough that some discussion is certainly in order.

One way of understanding the physics is to realize that ``dressing'' a state
does not commute with evolution through a time-dependent potential (here, the
gravitational collapse).

The in-vacuum, as we have seen, can be thought of as a bare vacuum ``dressed''
by the contributions of interactions.  
This dressing was determined by the Hamiltonian $H_{\rm p}$
in the distant past, where the
state was specified, by the requirement that the state indeed be 
the one of lowest energy. 
By contrast, consider a physical out-state with a thermal distribution of
photons.  This state is, to zeroth order, the one described by Hawking. 
However, if we wish to dress this state and retain its physical interpretation,
we must do so on the basis of the Hamiltonian $H_{\rm f}$ in the distant future.
(The dressed thermal state could be defined by requiring that the KMS
condition with respect to $H_{\rm f}$ hold.)
This Hamiltonian differs substantially from $H_{\rm p}$, in that the
electromagnetic potential operators of the two are related by a Bogoliubov
transformation.

The bare in-state dressed by $H_{\rm p}$ will not evolve to the bare out-state
dressed by $H_{\rm f}$. To see this, let us consider the situation in more
detail.

The in-vacuum was modified by the addition of 
triples of (bare) quanta.  Each of these triples
consists of an electromagnetic quantum and a pair of charged quanta.  While the
spatial momenta of each triple sum to zero, contributions with arbitrarily high
individual momenta are present.
The field modes describing these triples, after passage through the collapsing
region, are distorted.  We here neglect the distortion of the charged-particle
modes (we are interested only in charged particles which are everywhere far
from the collapse region).  But the photon modes are very severely
distorted.  Each photon mode which passes through the collapse
region is exponentially red-shifted; also the photon field modes undergo a
squeezing transformation (mixing positive and negative frequencies) which alters
the particle numbers.  

This means that the field modes in the future cannot recombine undistorted to
dress the bare out-state as would be prescribed by $H_{\rm f}$.  The out-state
will thus have an ultra-high frequency content which differs from that 
of the bare out-state
dressed by $H_{\rm f}$:  there will be ultra-energetic excitations.
And these distortions will be due to missing inward-directed photons,
leaving outward-directed charged-particle pairs.

The discussion just given points up a subtlety in dealing with
interacting field theories --- one must be careful to distinguish between bare
and dressed quantities, and in general one must be careful to justify the
physical interpretations of the quantities.  In our case, we have worked out the
amplitude~(\ref{expval}), which in principle exists in full quantum
electrodynamics, to lowest non-trivial order.  We are thus justified in
identifying it as a physically significant quantity.  It is true that its
simple
interpretation in terms of particles is only valid insofar as we may identify
the operators ${\tilde\psi}^+_{\rm f}$, $\Phi ^-_{\rm f}$, $\psi ^+_{\rm f}$ as
annihilation and creation operators for particles, and this becomes a
non-trivial issue for very high energies.  However, the problem of defining
the particle-content is not really relevant here.  What matters is that the
amplitude unambiguously represents ultra-energetic excitations of the
charged fields.

In this connection, it is worth pointing out that the amplitude~(\ref{evamp})
we have computed would vanish in a stationary space--time (assuming the
positive-/negative-frequency decomposition of fields is with respect to Killing
vector defining the stationarity), for then the Bogoliubov coefficient $\beta$
would be zero.  Thus if one wishes to view the amplitude as a certain
three-point function, the difference of this function between the case of
gravitational collapse and the stationary case is a clear signal of the effect
of gravitational collapse on the physics.

Finally, we point out that similar results will hold for holes with angular
momentum and charge.  We really only used the Schwarzschild character of the
hole in two places:  when we assumed the Bogolibubov coefficients were those of
a Schwarzschild hole; and in estimating the magnitude of the momentum $l$ as
$\sim\omega _{\rm H}\exp +u/(4M)$.  However, it is well understood what these
quantities are in the more general, Kerr--Newman, case.  The Bogoliubov
coefficents were discussed by Hawking, and the momentum increases as $\sim
\omega _{\rm H}\exp +\kappa u$, where $\kappa$ is the surface gravity of the
hole.

\subsection{Other Processes}

We chose to compute the amplitude for pair production because it was simple both
technically and conceptually.  Technically, it was a first-order process in
which only one term contributed significantly.  Conceptually, it had a clear
interpretation, as the first-order contribution to an expectation-value which
should exist in full, non-linear quantum electrodynamics.

Many other processes could be investigated, and similar conclusions reached. 
However, the results are harder computationally and are also subject to further
interpretational complications (largely associated with renormalization).  It
should be clear, however, that not only production of ultra-energetic pairs, but
also the scattering of particles initially present to ultra-high energies is
possible.  (Some processes would simply be by virtual photon exchange with the
ultra-energetic pairs described here.)  There are also corrections to vacuum
polarization effects, where a virtual photon can pass through the collapsing
region.

We shall not attempt to discuss any of these here.  The main point of the
present paper is that the exponential increase of the energies quickly carries
us beyond the point where we can make reliable quantum-field-theoretic
computations.

\section{Consequences and Predictions}
\label{CP}

We have seen that when interactions are taken into account, quantum field
theory in curved space--time leads to a very different picture of a black hole
than the one drawn by Hawking.  Instead of an essentially classical object
accompanied by the emission of low-energy ($\sim\omega _{\rm H}$) thermal
quanta, we have a prediction of emissions at exponentially increasing energy
scales ($\sim \omega _{\rm H}\exp +u/(4M)$), where the $e$-folding time
($4M\simeq (M/M_\odot)\times 2.0\times 10^{-5}$ s, with $M_\odot$ the mass of
the Sun) is fairly short for known black-hole candidates.

This has consequences for the general theoretical picture (the link between
black holes and thermodynamics) and possibly for experiment as well.  Too, we
must divide the discussion into the {\em cis-Planckian} regime, where the
energies involved are below the Planck scale, and the Planckian regime.

\subsection{The Cis-Planckian Regime}

While the computations above have been done only for one quantum-electrodynamic
process and only to first order, it should be clear that the underlying
principle is more general:  the passage of virtual quanta through the
collapsing region distorts them significantly and so upsets the balance of
virtual processes, resulting in real  effects which become ultra-energetic. 
The energy scale for these increases as $\sim\omega _{\rm H}\exp +u/(4M)$. 

Typically, these effects at a given point are due to contributions from the
blue-shifted precursors of Hawking photons. Of all the photons that might be
visible at a point, Hawking photons should come from (roughly) that fraction of
the sky occupied by the incipient black hole.   This fraction will be
appreciable for observers within a few Schwarzschild radii of the hole, and
will fall off as the square of the distance from the hole at greater
distances.  Thus we may expect significant effects in a volume of space whose
radius is a few Schwarzschild radii, and possibly measurable effects some
distance further out.

One would like to know the rate of production of ultra-energetic particles.
Unfortunately, the exponential increase of the energies makes reliable
computations of this first difficult, and then impossible.  The first-order
quantum-electrodynamic computation done above would be expected to be valid for
energies $\lesssim 1$~MeV (beyond which higher-order effects, due to  further
real and virtual electron--positron pairs, would have to be considered).  Above
about $100$~MeV one would have to consider meson production and associated
strong-force physics; above about $1$~GeV there would be baryon production;
above about $100$ GeV, electroweak mixing.
And much beyond this, we are in a regime in which we have few experimental data
and theory is very much a matter of speculation.  

It should be pointed out that there are essentially two competing effects
determining how significant a particular ultra-energetic amplitude is.  On one
hand, the exponential compression of the modes takes place in a region of
space--time which is itself being exponentially compressed (its extent in
advanced time is of the order of the blue-shifted period), and this tends to
decrease the amplitude.  On the other, as the energies increase one has the
potential of more processes contributing to a given amplitude, as well (perhaps)
as the break-down of perturbation theory itself.  We have no good ways of
estimating these latter effects, and thus we have no good way of estimating how
much radiation is produced.  We cannot even say whether the total production
would tend to increase or decrease with $u$.  We do know, however, that the
characteristic energy scale increases exponentially.

While this difficulty in making quantitative predictions is a very real one, we
also have the exciting prospect of the neighborhood of a black hole being an
experimental laboratory in which energies well beyond those terrestrially
accessible will appear.  And while the quantitative estimates of the rate of
production are at present unavailable, it is worth noting that there would be
one general characteristic of the effects described here which would distinguish
them from other high-energy astrophysical effects in the neighborhood of the
hole.  The energies of the effects here would be set essentially by the
red-shift factor $v'(u)$, and not the distance from the hole
(except for
additional red-shift effects very near the hole).  (The {\em probabilities}
of the effects would fall with the distance, as noted above, but the energies
would not.)  By contrast, most highly energetic astrophysical effects near a
black hole are driven by loss of potential energy, and so the energies involved
vary inversely with the distance from the hole.

As noted above, the exponential increase in the energy scales means that this
regime of cis-Planckian physics is run through in short order, typically a few
dozen $e$-foldings.   One might think that this means that only if we catch a
black hole on the verge of formation is this cis-Planckian regime accessible. 
This is not the case, however.  What matters is the increasing blue-shift of
the frequencies, and such an increase  can occur in other ways.  For example,
it may occur for an observer moving relative to an established black hole, if
as time passes in the observer's frame there is a direction in which there are
null geodesics passing closer and closer to the hole. It may also occur if
matter accretes to an existing black hole~\cite{Helfer:2001}.

It is worth noting that besides providing a window on physics at very high
energies, the effects here would provide some constraints on very low
energies.  This is because they rely, like the original Hawking computation, on
the assumption that there is an effectively massless field (the photon, in the
work above).   Here ``effectively'' massless means much lower in mass than the
scale $\omega _{\rm H}$.  Since the photon's mass is presently believed
constrained to be $<6\times 10^{-17}$~eV \cite{Eidelman:2004} and $\omega _{\rm
H}\simeq 5.3\times 10^{-12} (M_\odot /M)$~eV (where $M_\odot$ is the mass of
the Sun), a non-zero photon mass could cause a difference in the production of 
ultra-energetic effects by black holes of masses $\gtrsim 10^5 M_\odot$. 
Finally, in this connection we note that neutrinos of masses below $\omega
_{\rm H}$ could also contribute to ultra-energetic effects in manners similar
to that of the photon; current experiments only constrain the mass of the
lightest neutrino to be $\lesssim 10^{-2}$~eV.

\subsection{The Planck Regime}

The most interesting feature of the analysis here is that it unequivocally
demonstrates the inadequacy of conventional quantum theory in curved
space--time to describe physics in the neighborhood of a black hole.  That
conventional theory would predict the existence of effects at energies
increasing exponentially beyond the Planck scale, and at that point the neglect
of quantum-gravitational effects is illegitimate.   Without a theory of quantum
gravity, we cannot say how this conventional model breaks down --- whether it
is conventional quantum theory which becomes inadequate, or the treatment of
space--time by a classical model, or both --- but we do
know that one of these elements must be altered.  

In particular, we cannot say whether an established black hole (one for which
the energy scale in question has entered the Planck regime) will emit quanta at
all, in any frequency range.  We cannot even say whether the quantum structure
of the hole and its neighborhood
will be effectively stationary or not.  Because of the appearance of the
Planck scale, the problem is wholly dependent on quantum-gravitational
physics, not presently understood.

The simplest hypothesis would be that, whatever the details of the
quantum-gravitational structure are, they result in the emission of particles
at energies somewhat below the Planck energy.  (Of course, without a theory of
quantum gravity, we could say nothing about the production rates for different
species.) If this is correct, and the cut-off is independent of position, then,
as in the cis-Planckian case, the limiting energy would be essentially
independent of distance from the hole (apart from possible red-shift effects
for particles very near the hole), although the  probability of emission would
fall with distance from the hole. Again, this would be a signature likely to
distinguish these effects from other ultra-energetic astrophysical effects near
the hole.  And again, with our present ignorance of the physics of the extreme
energy scales involved, we can only speculate about the sorts of particles that
might be produced.  It is possible that this mechanism is responsible for the
production of ultra-high energy cosmic rays.  

For established black holes, besides the  possibility of direct observation of 
particle emission,  there are other potentially significant astrophysical
consequences. Whatever hypotheses on the rate of particle production are made,
the rate of mass loss for the hole is likely to be different from
that predicted by Hawking, leading to a different black-hole
lifetime. These differences could have significance for cosmological models,
since density fluctuations in the early Universe may have produced black
holes.  Up to now, the main constraints on the number of these have been due to
the cosmological implications of Hawking radiation.  A difference in the
predicted radiation would alter these constraints.

\subsection{Theoretical Implications}

The results here, besides showing the necessity of developing some aspects of a
theory of quantum gravity for the treatment of quantum fields near black holes,
imply that we must reconsider the picture that has been broadly accepted of
quantum theory and black-hole thermodynamics.

Let us recall that, before the theory of black-hole radiance, a strong formal
parallel had been noted between the theory of black holes and thermodynamics. 
The zeroth law (the existence of a well-defined temperature) was the fact that
the surface gravity of a stationary black hole was constant over the horizon;
the first law was conservation of energy; and the second was that the area of a
black hole could only increase.  Thus it seemed that one should interpret the
surface gravity and area of a black hole as a sort of temperature and entropy,
respectively.

The relation remained only formal, however, until an explicit link with
ordinary thermodynamics could be found.  Such a link was first proposed by
Bekenstein~\cite{Bekenstein:1973}, based on the notion of information loss. 
This argument was however difficult to make completely quantitatively precise. 
Then Hawking predicted that black holes actually radiate thermally with a
well-defined temperature $\omega _{\rm H}$ (which turns out to be the surface
gravity over $2\pi$).  This seemed to explain one half of the puzzle, that of
giving a thermodynamic meaning to the black-hole temperature.  It was also true
that from that result and the thermodynamic relation $dQ=TdS$ a definite value
for the black-hole entropy could be inferred, but the problem of providing a
convincing independent thermodynamic interpretation for black-hole entropy
remained open.  And despite much ingenious work (e.g., attempts to count the
number of black-hole states quantum-gravitationally, thought-experiments based
on having ordinary thermal systems interact with black holes), the problem is
still open~\cite{Helfer:2003}.

The results here call into question the prediction of black-hole radiation, and
with that the explanation of black-hole temperature.  They do not, as
emphasized above, mean that black holes do not radiate, but they do make the
previous analysis untenable.    Their real lesson is that the regime in
question cannot be understood at all without quantum gravity.

\subsection{Conclusions}

We may thus summarize our picture as follows.  The black hole has an
essentially quantum character, with significant ultra-energetic quantum effects
in a region extending on the order of a Schwarzschild radius beyond the hole. 
As the hole forms, the energy scales in question increase exponentially, making
the neighborhood of the incipient hole a potential laboratory for studying
quantum fields at all energy scales.  (Such increasing energy scales may also
be obtained for established holes, for observers moving relative to them or
when matter accretes onto them.)  After a finite number of $e$-foldings,
however, the  energy scales have reached the Planck regime, and we are at the
limit of known physics. At this point, the neighborhood of the hole has an
essentially quantum-gravitational structure.  We may hope that studies of black
holes may bring experimental evidence for the character of this structure.


\section*{Appendix:  Previous Work on Interactions}

The main argument of this paper has been that the effects of interactions among
quantum fields substantially alter their physics in a gravitationally
collapsing space--time.  This differs from most previous
thinking; I shall here briefly discuss the relation of the ideas here to
previous work on interactions.

Probably the bulk of the work that has been done has grown out of attempts to
model Hawking-radiating mini black holes whose temperatures are high enough to
make them astrophysically interesting.  However, to my knowledge, all such work
is based on the {\em assumption} that the black hole acts as a black (or brown)
body of a given size and temperature.  Thus this approach {\em assumes} that a
Hawking-type result applies even for interacting fields; it does not attempt to
justify this result.  

There has  also been the suggestion, not fully developed, that general
principles of quantum field theory and invariance should force a black hole to
radiate thermally even when quantum-theoretic nonlinearities are
present~\cite{Wald:1994}.  The main support for this notion comes from  the
Bisognano--Wichmann Theorem, which can be interpreted as stating that a
uniformly accelerating observer in Minkowski space would perceive the
$n$--point functions of the vacuum of even an interacting field theory to be
the same as those of a thermal state at the corresponding Unruh temperature.
However, it has so far not been possible to develop a theorem like this in the
gravitational-collapse case.  (The Bisognano--Wichmann Theorem is very much a
result of {\em special}-relativistic field theory, and in particular depends
crucially on the existence of a semi-bounded self-adjoint Hamiltonian, a
hypothesis which is invalid in the gravitational-collapse
region~\cite{Helfer:1996}.)   

Perhaps the best attempt to confront the problem of interactions has been due
to Gibbons and Perry~\cite{GP:1976}.  These authors  outlined an argument that
the Feynman propagators for the theory would have a periodicity in imaginary
time which would enforce thermality of the state even in the case of
interactions.  However, there are some subtleties which make it difficult to
completely prosecute the argument.  The details of the development of the
theory by Feynman diagrams, and especially the question of the choice of
dressing of the in- and out-states, were not spelled out.  Also, while one of
the authors' main points was the necessity of showing that an initially vacuum
state would dynamically equilibrate to a thermal one, in fact it is
unrealistically hard to follow this dynamical process, and so they based the
details of their analysis on the use of the Hartle--Hawking
propagator~\cite{HH:1976}.   This is a somewhat formal object usually
interpreted as representing a black hole {\em assumed} in equilibrium with a
thermal bath of radiation.    It is an object which {\em by construction} has
no explicit information about a collapse phase; it  was derived, and is most
often used, under the assumption that  one is dealing with late-time physics
for which neglect of the collapse  phase is legitimate.  Thus the
Gibbons--Perry analysis would not uncover the sorts of effects investigated
here, which turn on the propagation of virtual quanta from before the time of
collapse through the collapse region.

\end{document}